\documentclass[11pt,a4paper,svgnames]{article}
\pdfoutput=1

\usepackage[utf8]{inputenc}
\usepackage{uniinput}
\usepackage{jheppubmod}
\usepackage{hyperref}
\usepackage{amsmath,mathtools,bbm}
\usepackage{lipsum}
\usepackage[plain]{fancyref}

\allowdisplaybreaks

\renewcommand{\d}{\mathrm{d}}
\newcommand{\ie}{\textit{i.e.~}}
\newcommand{\eg}{\textit{e.g.~}}

\usepackage{relsize}
\newcommand{\s}[1]{{\mathsmaller{#1}}}

\usepackage{microtype}

\title{Black holes with spindles at the horizon}
\author[a,b]{Suvendu Giri~}
\emailAdd{suvendu.giri@unimib.it}
\affiliation[a]{Dipartimento di Fisica, Università di Milano-Bicocca, I-20126 Milano, Italy}
\affiliation[b]{INFN, sezione di Milano-Bicocca, I-20126 Milano, Italy}

\abstract{We construct $\mathrm{AdS}_4 \times Σ$ and $\mathrm{AdS}_2 \times Σ \times Σ_\mathfrak{g}$ solutions in F(4) gauged supergravity in six dimensions, where $Σ$ is a two dimensional manifold of non-constant curvature with conical singularities at its two poles, called a spindle, and $ Σ_\mathfrak{g}$ is a constant curvature Riemann surface of genus $\mathfrak{g}$. We find that the first solution realizes a ``topologically topological twist'', while the second class of solutions gives rise to an ``anti twist''.
We compute the holographic free energy of the $\mathrm{AdS}_4 \times Σ$ solution and find that it matches the entropy computed by extremizing an entropy functional that is constructed by gluing gravitational blocks. For the $\mathrm{AdS}_2 \times Σ \times Σ_\mathfrak{g}$ solution, we find that the Bekenstein-Hawking entropy is reproduced by extremizing an appropriately defined entropy functional, which leads us to conjecture that this solution is dual to a three dimensional SCFT on a spindle. 
A class of the $\mathrm{AdS}_2 \times Σ \times Σ_\mathfrak{g}$  solutions can be embedded in four dimensional $T^3$ gauged supergravity, which is a subtruncation of the six dimensional theory.
}
\begin{document}
\maketitle

\section{Introduction and summary}
An important tool to study the AdS/CFT correspondence has been to construct supersymmetric solutions by wrapping branes on supersymmetric cycles.
 was originally done by following the idea of \cite{Maldacena:2000mw}, where
in order for the lower dimensional solution to preserve supersymmetry, the background R-symmetry gauge field cancels the spin connection on the compactification manifold -- a mechanism dubbed ``topological twisting''. The holographic duals of these gravity theories are topologically twisted superconformal field theories (SCFTs) \cite{Bershadsky:1995qy}. Such constructions with branes wrapped on a two dimensional constant curvature Riemann surface $Σ_\mathfrak{g}$ of genus $\mathfrak{g}$ have been extensively studied \eg, for M2 branes in \cite{Cacciatori:2009iz}, M5 branes in \cite{Bah:2012dg}, D3 branes in \cite{Benini:2013cda}, and D4 branes in \cite{Bah:2018lyv}.

Recently, new AdS/CFT constructions which do not rely on topological twisting have been studied, starting with \cite{Ferrero:2020laf} where D3 branes wrapped on a two dimensional surface known as a spindle were studied. The spindle is a weighted projective plane $\mathbb{WCP}^1_{[n_\pm]}$, with weights given by two positive coprime integers $n_\pm$. The metric on such a space is regular everywhere except at the poles where there are conical singularities parametrized by the integers $n_\pm$. This compactification preserves supersymmetry, not via a topological twist, but rather via an ``anti twist'' \cite{Ferrero:2020laf} or  a ``topologically topological twist'' \cite{Ferrero:2021wvk}.
This has paved the way for a new class of constructions where branes are wrapped on a spindle instead of a constant curvature Riemann surface.
M5 and M2 branes were studied in \cite{Ferrero:2021wvk} and \cite{Ferrero:2020twa}, a multicharge solution from D3 branes was studied in \cite{Hosseini:2021fge}, and very recently, D4 branes were studied in \cite{Faedo:2021nub}.
A family of charged rotating solutions of the form AdS$_2 × Σ$ were constructed in 4d $\mathcal{N}=4$ gauged supergravity in \cite{Ferrero:2021ovq}.
Compactifications on a topological disc, which preserve supersymmetry in a similar way, have been studied \eg in \cite{Bah:2021hei,Bah:2021mzw}. It was shown in  \cite{Couzens:2021tnv} that D3 branes wrapped on a topological disc are a different global completion of the same local solution as in \cite{Boido:2021szx}. Supersymmetric solutions corresponding to D3 branes and M5 branes wrapped on a topological  disc were constructed in \cite{Suh:2021ifj},  D4-D8 solutions were constructed in \cite{Suh:2021aik}, and M2 branes in \cite{Suh:2021hef}.

In this paper, we construct a new class of solutions of the form $\mathrm{AdS}_2 \times Σ \times Σ_\mathfrak{g}$, starting from six dimensional F(4) gauged supergravity. This arises as a consistent truncation of massive IIA (mIIA) supergravity compactified on a warped $S^4$. We find that these solutions preserve supersymmetry via an ``anti twist''.
These solutions can be interpreted in the four dimensional theory obtained by compactifying the six dimensional theory on $Σ_\mathfrak{g}$. One such class of solutions exists for a specific choice of parameters and corresponds to the ``gauged $T^3$ supergravity'', while a second class corresponds to minimal gauged supergravity in four dimensions.
These four dimensional theories can also be uplifted to eleven dimensional supergravity on $\mathrm{AdS}_4 \times S^7$, which is dual to three dimensional ABJM theory \cite{Aharony:2008ug}. So the four dimensional gravity solutions can also be related to the ABJM theory.
We compute the entropy by extremizing an off-shell entropy functional obtained by appropriately gluing ``gravitational blocks'' \cite{Hosseini:2019iad}. Remarkably we find that the result agrees with the computation in gravity.
We also construct supersymmetric $\mathrm{AdS}_4 \times Σ$ solutions in the six dimensional F(4) gauged supergravity, where we instead find that supersymmetry is preserved via ``a topologically topological twist''. We expect this to be dual to a five dimensional SCFT on the spindle. We again compute the free energy on $S^3$ by extremizing the entropy functional constructed by gluing gravitational blocks and find agreement with the gravity result obtained from our solution.

The outline of this paper is as follows. We begin with a quick overview of the six dimensional F(4) gauged supergravity theory in \fref{sec:F4}. We then construct the supersymmetric $\mathrm{AdS}_4 \times Σ$ solution in \fref{sec:ads4}, and the supersymmetric $\mathrm{AdS}_2 \times Σ \times Σ_\mathfrak{g}$ solutions in \fref{sec:ads2}. We conclude with some discussion in \fref{sec:discussion}.

\vskip 10pt
\noindent
\textbf{Note:} While writing up this paper, \cite{Faedo:2021nub} appeared on \texttt{arXiv} whose results partially overlap with ours. They construct $\mathrm{AdS}_4 \times Σ$ solutions in 6d F(4) gauged supergravity and consistent with our observations, they also find that the solution is realized as ``a topologically topological twist''.

\section{6d F(4) gauged supergravity}\label{sec:F4}
We begin by recalling some important aspects of six dimensional F(4) gauged supergravity.
F(4) superalgebra is the minimal extension of the SO(2,5) symmetry group of six dimensional AdS. It contains $\mathfrak{so}(2,5) \otimes \mathfrak{su}(2)$ as the maximal bosonic subalgebra, and is therefore the natural candidate for a six dimensional supergravity theory with $16$ supercharges \ie, $\mathcal{N}=2$ in $d=6$. The minimal F(4) supergravity theory (containing only the gravitino multiplet) was constructed in \cite{Romans:1985tw}, while the theory coupled to vector multiplets (which are the only possible massive long multiplets in $\mathcal{N}=2$) was constructed in \cite{DAuria:2000afl,Andrianopoli:2001rs}.

The bosonic fields contained in the gravitino multiplet are the metric $g_{μν}$, four gauge fields $A^α_μ$ corresponding to the symmetry group U(1)$\times$ SU(2)$_\s{R}$ (where $α \in \{0,r\}$, with $r \in \{1,2,3\}$ being an index in the adjoint representation of SU(2)$_\s{R}$), a two form $B_{μν}$ and the dilaton $σ$, where $μ,ν \in\{ 0,1…,5\}$ are spacetime indices. The fermionic fields consist of two gravitini $ψ^A_μ$, and two spin-$1/2$ fermions $χ^A$, where $A \in \{1,2\}$, transforming in the fundamental representation of SU(2)$_\s{R}$.

The gravity multiplet can be coupled to $n_\s{V}$ vector multiplets labelled by an index $I \in \{1,…,n_\s{V}\}$. Each vector multiplet contains a gauge field $A_μ$, four scalars $ϕ_α$, and a spin-$1/2$ fermion $λ_A$, where $α$ and $A$ are indices in the adjoint and fundamental representations of SU(2)$_\s{R}$ respectively, as above. The $4n_\s{V}$ scalars span the coset manifold
\[ \frac{\mathrm{SO}(4,n_\s{V})}{\mathrm{SO}(4) \times \mathrm{SO}(n_\s{V}) }\, . \]
The scalars $ϕ_α$ can be encoded in a coset representative $L^Λ{}_Σ \in \mathrm{SO}(4,n_\s{V})$, where $Λ \in \{ α,I\}$ and $I$ counts the number of vector multiplets $I \in \{ 1,…,n_\s{V} \}$. The gauged six dimensional theory can be obtained by a compact gauging of $\mathcal{G} = \mathrm{SU}(2)_\s{R} \times G$, where $G$ is a $n_\s{V}$ dimensional compact subgroup of SO$(n_\s{V})$.
The six dimensional bosonic Lagrangian, in the notation of \cite{Andrianopoli:2001rs} is:
\begin{equation}\label{eq:L}
\begin{split}
	\mathcal{L} = &-\frac{R}{4} - \frac{1}{8}e^{-2σ}\mathcal{N}_{ΛΣ}\hat{F}^Λ_{μν}\hat{F}^{Σμν} + \frac{3}{64}e^{4σ} H_{μνρ}H^{μνρ} + ∂^μσ∂_μσ - \frac{1}{4}P^{Iαμ} P_{Iαμ}\\
	 &- \frac{1}{64}ϵ^{μνρσλτ} B_{μν} 
	\left( η_{ΛΣ} \hat{F}^Λ_{ρσ}\hat{F}^{Σ}_{λτ} + m B_{ρσ} \hat{F}^0_{λτ} + \frac{1}{3}m²B_{ρσ}B_{λτ} \right)  - V\, ,
\end{split}
\end{equation}
where
\begin{equation}\label{eq:calNP}
\begin{split}
	\mathcal{N}_{ΛΣ} &= L_Λ{}^α \left(L^{-1}\right)_{αΣ} - L_Λ{}^I \left(L^{-1}\right)_{IΣ}\,,\\
	P^I_α &= \left(L^{-1}\right)^I{}_Λ \left( \d L^Λ{}_α - f_Γ{}^Λ{}_Π A^ΓL^Π{}_α \right)\, ,
\end{split}
\end{equation}
with $f^Λ{}_{ΠΓ}$ being the structure constants of the gauge group $\mathcal{G}$. $g$ is the gauge coupling constant and $m$ is the mass parameter associated to the two form. 
The minimal six dimensional F(4) gauged supergravity theory can be obtained as a consistent truncation of massive type IIA supergravity in ten dimensions on a warped S$^4$ \cite{Cvetic:1999un}. This was dualized to a truncation of type IIB supergravity via a non-Abelian T-duality in \cite{Jeong:2013jfc}, and was further generalized to a large class of geometries in \cite{Hong:2018amk}.
Substantial evidence was provided in \cite{Hosseini:2018usu} that even the six dimensional theory coupled to a vector multiplet can be obtained as a consistent truncation of ten dimensional mIIA supergravity. It was also shown in \cite{Malek:2019ucd} that the theory with one vector multiplet can be obtained from a consistent truncation of type IIB supergravity on a general class of manifolds, which includes the Abelian T-dual of the mIIA background considered here.
The parameter $m$ is then related to Romans' mass $m=F_{(0)}$.

The field strength  
$ \hat{F}^Λ_{ρσ} = F^Λ_{ρσ} - m δ^{Λ0} B_{μν} $
is dressed with this mass parameter, and we use the non-standard convention of \cite{Andrianopoli:2001rs} where $F = F_{μν} \d x^μ∧\d x^ν$, with $F_{μν} = \left(∂_μ A_ν - ∂_ν A_μ\right)/2$.
Variation of the fermions (upto linear order in fermions) under an infinitesimal supersymmetry transformation are given by
\begin{equation}\label{eq:bpseqs0}
\begin{split}
	δψ_{Aμ} &= ∇_μϵ_A - \frac{i}{2}g σ^r_{AB} A_{rμ}ϵ^B + \frac{1}{16}e^{-σ} \left[ \hat{T}_{[AB]νλ}Γ - T_{(AB)νλ} \right] \left( Γ_μ{}^{νλ} - 6δ^ν_μ Γ^λ \right)ϵ^B\\
	&\quad + \frac{i}{32}e^{2σ}H_{νλρ} Γ \left( Γ_μ{}^{νλρ} - 3δ^ν_μ Γ^{λρ} \right)ϵ_A + S_{AB}Γ_μ ϵ^B\,,\\
	δχ_A &= \frac{i}{2}Γ^μ ∂_μσ ϵ_A + \frac{i}{16}e^{-σ} \left[ \hat{T}_{[AB]νλ}Γ + T_{(AB)νλ} \right]Γ^{νλ}ϵ^B + \frac{1}{32}e^{2σ} H_{νλρ} Γ Γ^{νλρ}ϵ_A + N_{AB}ϵ^B\,,\\
	δλ^I_A &= i P^I_{rμ} σ^r_{AB} Γ^μ ϵ^B - iP^I_{0μ}ϵ_{AB}ΓΓ^μϵ^B + \frac{i}{2}e^{-σ}T^I_{μν}Γ^{μν}ϵ_A + M^I_{AB}ϵ^B\,,
\end{split}
\end{equation}
where $Γ$ is the six dimensional chirality matrix $Γ=iΓ^0…Γ^5$, and the dressed vector field strengths are defined as
\begin{equation}
	\hat{T}_{[AB]νλ} = ϵ_{AB}\left(L^{-1}\right)_{0Λ}\hat{F}^Λ_{νλ}\,, \quad
	T_{(AB)νλ} = σ^r_{AB} \left(L^{-1}\right)_{rΛ}F^Λ_{νλ}\,, \quad
	T_{Iνλ} = \left(L^{-1}\right)_{IΛ}F^Λ_{νλ}\,,
\end{equation}
and $S_{AB}, N_{AB}, M_{AB}$ represent the extra contributions to the fermion variations due to gauging and the mass parameter. Greek indices are raised and lowered with the SO(4,$n_\s{V}$) invariant matrix $η_{ΛΣ} = \mathrm{diag}\left(1,1,1,1,-1,\ldots,-1\right)$, and Roman indices with the SU(2)$_\mathrm{R}$ tensor $ϵ_{AB}$

We further restrict ourselves to a theory which contains only one vector multiplet $n_\s{V}=1$. We can consistently set all gauge fields to zero except $A^{r=3}_μ$ and $A^{I=1}_μ$ which will be necessary for the twisting, and for providing a magnetic charge for the black hole. Additionally, we require that the scalar fields in the vector multiplet are singlets under the gauge field $A^{r=3}_μ$. Furthermore, requiring that the black holes are purely magnetic restricts the only non-zero component of the scalars to be $ϕ_3$. Following \cite{Hosseini:2018usu,Suh:2018szn}, we choose a convenient parametrization of the scalar coset given by
\begin{equation}
	L^Λ_Σ = \begin{pmatrix}
		1 & 0 & 0 & 0 & 0\\
		0 & 1 & 0 & 0 & 0\\
		0 & 0 & 1 & 0 & 0\\
		0 & 0 & 0 & \cosh ϕ_3 & \sinh ϕ_3\\
		0 & 0 & 0 & \sinh ϕ_3 & \cosh ϕ_3
		\end{pmatrix}\,.
\end{equation}
With this parametrization, the kinetic matrix for the vector fields follows from \fref{eq:calNP}
\begin{equation}
	N_{ΛΣ} = \begin{pmatrix}
		1 & 0 & 0 & 0 & 0\\
		0 & 1 & 0 & 0 & 0\\
		0 & 0 & 1 & 0 & 0\\
		0 & 0 & 0 & \cosh 2ϕ_3 & -\sinh 2ϕ_3\\
		0 & 0 & 0 & -\sinh 2ϕ_3 & \cosh 2ϕ_3
		\end{pmatrix}\,.
\end{equation}
In this parametrization, 
the shifts $S_{AB}, N_{AB}, M_{AB}$ become
\begin{equation}
\begin{split}
	S_{AB} &= \frac{i}{4}\left(g e^σ \cosh ϕ_3 + m e^{-3σ}\right)ϵ_{AB}\,,\\
	N_{AB} &= \frac{1}{4}\left(g e^σ \cosh ϕ_3 - 3m e^{-3σ}\right)ϵ_{AB}\,,\\
	M_{AB} &= -2g e^σ \sinh ϕ_3\, σ^3_{AB}\,.
\end{split}
\end{equation}
The 6d theory has $\mathrm{AdS}_6$ as a vacuum solution if $g=3m$. We are interested in near horizon solutions of higher dimensional objects whose full solution would represent a flow from $\mathrm{AdS}_6$ to $\mathrm{AdS}_4 \times Σ$ or $\mathrm{AdS}_2 \times Σ \times Σ_\mathfrak{g}$ respectively. So we will choose $g=3m$ in the rest of paper.

\section{The supersymmetric $\mathrm{AdS}_4 \times Σ$ solution}\label{sec:ads4}
\subsection{Supersymmetry equations}
We are interested in a solution of the form $\mathrm{AdS}_4 \times Σ$. To find this, we consider the following ansatz for the metric
\begin{equation}\label{eq:ads4ansatz}
	\d s² = w(y) \left[\frac{4}{9}\d s²_{\mathrm{AdS}_4} - \frac{\d y²}{q(y)} - \frac{q(y)}{r(y)}\d z² \right] \,,
\end{equation}
and assume that the two-form vanishes \ie, $B_{μν}=0$.
Let us pick the non-zero components of the two gauge fields to lie only along the spindle $A^{3,I}=A^{3,I}_z(y) \d z$, where the index $I$ labels the gauge field from the vector multiplet. Maxwell's equations follow from the Lagrangian in \fref{eq:L}, and in this case read
\begin{equation}
	∂_y \left(\sqrt{-g}\,e^{-2σ} \mathcal{N}_{ΛΣ}F^Σ_{yz}g^{yy}g^{zz}\right)=0\,.
\end{equation}
This determines the gauge field strengths along the spindle\footnote{For brevity, we will often not write the $y$ dependence of the functions explicitly.}
\begin{equation}\label{eq:F3FI-6d}
\begin{split}
	F^3 = -\frac{e^{2σ}}{w}\left(f_3\cosh2ϕ_3+f_i\sinh 2ϕ_3\right) \mathrm{vol}_Σ\,,\quad
	F^I = -\frac{e^{2σ}}{w}\left(f_3\sinh 2ϕ_3+f_i\cosh2ϕ_3\right) \mathrm{vol}_Σ\,,
\end{split}
\end{equation}
where $f_3, f_i$ are constants.
We choose the following representation for the gamma matrices
\begin{equation}
	Γ^a = γ^a \otimes σ_3,\quad Γ^4 = \mathbbm{1} \otimes i σ_2, \quad Γ^5 = \mathbbm{1} \otimes i σ_1\,,
\end{equation}
where $a \in \{ 0,1,2,3\}$ are frame indices along the AdS$_4$ and indices $4,5$ are frame indices along the spindle. 
In six Lorenzian dimensions, spinors form a symplectic-Majorana pair which transform as $\mathcal{B}_6 ϵ_A = ε^{AB} ϵ_B^*$ under the six dimensional matrix $\mathcal{B}_6$ defined by $\mathcal{B}_6 Γ_m \mathcal{B}_6^{-1} = -Γ_m^*$, where $m \in \{0,…,5\}$. We choose $\mathcal{B}_6=\mathcal{B}_4 \otimes \mathcal{B}_2$, where $\mathcal{B}_4$ and $\mathcal{B}_2$ are matrices in 4d Lorenzian and 2d Euclidean space defined by $\mathcal{B}_4 γ_a \mathcal{B}_4^{-1}=γ_a^*$, and $\mathcal{B}_2 γ_i \mathcal{B}_2^{-1} = -γ_i^*$ respectively. $γ_a$ and $γ_i = \left(i σ_2, i σ_1\right)$, are the 4d Lorenzian and 2d Euclidean gamma matrices respectively. In particular, $\mathcal{B}_2$ is proportional to $σ_1$. With this choice, $\mathcal{B}_4 \mathcal{B}_4^*=-1$, $\mathcal{B}_2 \mathcal{B}_2^* = 1$ and $\mathcal{B}_6 \mathcal{B}_6^*=-1$.

We make an ansatz for the 6d spinor to be of the form $ϵ_1 = β_- \otimes η_1$, and $ϵ_2 = β_+ \otimes η_2$, 
where $β_\pm$ satisfy $∇_a β_\pm = \pm (i/2)γ_a β_\pm $. For the choice of $\mathcal{B}_4$ above, 
$\mathcal{B}_4 β_\pm = \mp \left(β_\mp\right)^*$. The action of $\mathcal{B}_2$ is given by $\mathcal{B}_2 η_1 = η_2^*, \mathcal{B}_2 η_2 = η_1^*$. In our conventions, $-ε_{AB} ϵ^B = ϵ_A$, $(σ^3)^A{}_B$ is the usual third Pauli Matrix and $σ^3_{AB} = -ε_{AC}\left(σ^3\right)^C{}_B$. Furthermore, we choose a gauge where the spinor is independent of the coordinate $z$.
	
We are looking for a supersymmetric solution. For simplicity, we pick a purely bosonic background by setting all the fermions to zero, and further demand that they remain zero under a supersymmetry transformation. This is imposed by demanding that the fermionic variations in \fref{eq:bpseqs0} vanish.
For the gravitino variation, this implies:
\begin{equation}
	\label{eq:δψ12new}
	\begin{split}
		δψ_{1a} &= γ_a β_- \otimes \left( -η_{1} - A η_{1} - B σ_3 η_{1} - \frac{iw^\prime \sqrt{q}}{3w}σ_1 η_{1} \right)\,,\\ 
		δψ_{2a} &= γ_a β_+ \otimes \left( η_{2} + A η_{2} - B σ_3 η_{2} - \frac{iw^\prime \sqrt{q}}{3w}σ_1 η_{2} \right)\,.
	\end{split}
\end{equation}
where
\begin{equation}
	A \coloneqq \frac{e^σ}{6w^{3/2}} \left(f_3\cosh ϕ_3 + f_i \sinh ϕ_3\right),\quad B \coloneqq \frac{√w}{3}\left(ge^σ\cosh ϕ_3 + me^{-3σ} \right).
\end{equation}
The equations for $η_{1,2}$ obtained above are invariant under the symplectic-Majorana condition, as expected. Since the two spinors $η_1$ and $η_2$ are related by $η_1 = \mathcal{B}_2 η_2^*$, there is only one independent 2d spinor, and all the equations can be written in terms of it. Relabelling $η_2$ as $ξ$, we can now rewrite the set of equations in \eqref{eq:bpseqs0} as:
\begin{equation}
\label{eq:KSE_AdS4}
\begin{split}
	δψ_{Aa} &: \left[1+\frac{e^σ}{6w^{3/2}}\left(f_3\cosh ϕ_3 + f_i \sinh ϕ_3 \right)\right] ξ - \frac{√w}{3}\left(ge^σ\cosh ϕ_3 + me^{-3σ} \right) σ_3 ξ\\
	&\quad - \frac{iw^\prime \sqrt{q}}{3w}σ_1 ξ \overset{!}{=} 0 \,,\\
	δψ_{A4} &: -\frac{3ie^σ}{8w²}\left(f_3\cosh ϕ_3 + f_i \sinh ϕ_3 \right)σ_1 ξ - \frac{1}{4}\left(ge^σ\cosh ϕ_3 + me^{-3σ} \right)σ_2 ξ 
	+ \sqrt{\frac{q}{w}}ξ^\prime \overset{!}{=} 0 \,,\\
	δψ_{A5} &: \frac{3ie^σ}{8w²}\left(f_3\cosh ϕ_3 + f_i \sinh ϕ_3 \right)σ_2 ξ - \frac{1}{4}\left(ge^σ\cosh ϕ_3 + me^{-3σ} \right)σ_1 ξ -\frac{ig}{2}\sqrt{\frac{r}{w q}}A^3_z ξ \\
	&\quad + i \frac{\sqrt{r}}{2w}\left(\sqrt{\frac{wq}{r}}\right)^\prime σ_3 ξ \overset{!}{=} 0 \,,\\
	δχ &: \frac{1}{2}\sqrt{\frac{q}{w}} σ^\prime σ_2 ξ - \frac{e^σ}{8w²}\left(f_3\cosh ϕ_3 + f_i \sinh ϕ_3 \right)σ_3 ξ + \frac{1}{4}\left(ge^σ\cosh ϕ_3 -3 me^{-3σ} \right)ξ \overset{!}{=} 0 \,,\\
	δλ &: \sqrt{\frac{q}{w}} ϕ_3^\prime σ_2 ξ - \frac{e^σ}{w²}\left(f_3\sinh ϕ_3 + f_i \cosh ϕ_3 \right)σ_3 ξ + 2g\sinh ϕ_3 e^σ ξ \overset{!}{=} 0 \,.
\end{split}
\end{equation}
\subsection{The solution}
We solve the first equation by choosing $q(y)=q_1(y) q_2(y)$, and $ξ=n(y)\left(\sqrt{q_1(y)},i\sqrt{q_2(y)}\right)$. This determines $q_1,q_2$ to be
\begin{equation}
\label{eq:q1q2_AdS4}
	q_{1,2} = \frac{3w}{w^\prime} \left[\pm \left(1 + \frac{e^σ}{6w^{3/2}}\left(f_3\cosh ϕ_3 + f_i \sinh ϕ_3 \right)\right) + \frac{√w}{3}\left(ge^σ\cosh ϕ_3 + me^{-3σ} \right)\right]\,.
\end{equation}
With this choice of the spinor $ϵ$, the four components of the last two equations immediately simplify to just two linearly independent equations which are solved by
\begin{equation}
	e^{4σ} = \cosh ϕ_3 + \frac{f_3}{f_i}\sinh ϕ_3\,,\quad
	w = \frac{\left(2f_i\, \mathrm{csch}\, ϕ_3\right)^{2/3}}{3^{4/3}}\left(\cosh ϕ_3 + \frac{f_3}{f_i}\sinh ϕ_3\right)^{1/6}\,.
\end{equation}
Inserting these in the gravitino variation along $Σ$ gives a full solution in terms of the scalar field $ϕ_3$. 
However, is convenient to choose a parametrization $ϕ_3 = \mathrm{arccoth}(y)$ to further simplify the solution. The full solution including the gauge field, the scalars, and the normalization of the spinor, in this parametrization is as follows:
\begin{equation}
\begin{split}
	w &= \frac{2^{2/3}}{3^{4/3}}\sqrt{f_i}\left(f_3+f_i \,y\right)^{1/6} \left(y²-1\right)^{1/4}\,,\quad
	r = r_0 \left(f_3 + f_i \,y\right)^{2/3} \left(y²-1\right)\,,\\
	q_{1,2} &= \pm \left( \frac{9f_3}{2f_i} + \frac{9y}{2} \right) + 2\cdot 6^{1/3} m \left(f_3 + f_i \,y\right)^{1/3}\sqrt{y²-1}\,,\\
	A^3_z &= \frac{6^{1/3}\left(3f_3\, y + 2f_i + f_i \,y²\right)}{f_i \left(y²-1\right)}\,, \quad
	e^{4σ} = \frac{f_3 + f_i \,y}{f_i \sqrt{y²-1}}\,,\\
	n &= n_0\left(f_3+f_i \,y\right)^{-1/8} \left(y²-1\right)^{-3/16}\,.
\end{split} 
\end{equation}
\subsection{Regularity of the solution}
For the metric to have a definite signature, the functions $q$ and $r$ appearing in the metric must be positive throughout the interval on which it is defined. Taking $y>0$, the metric function $r$ is positive for $y>1$, while the function $q=q_1 q_2$ is a polynomial of degree 8.
We want to find conditions for which the metric
\begin{equation}\label{eq:metricspindle}
	\d s²_Σ = \frac{\d y²}{q}+\frac{q}{r}\d z²\,,
\end{equation}
is a smooth metric on the spindle \ie, a two dimensional weighted projective plane with weights represented by two positive coprime integers $n_\pm$: $\mathbb{WCP}^1_{[n_+,n_-]}$. For this, $q$ must be positive in an interval $y\in[y_1,y_2]$ where $q\left(y_1\right)=q\left(y_2\right)=0$, and $y_2>y_1>0$.\footnote{\label{fn:positiveroots} In this article, we have chosen the roots to be positive. This is a choice and not a requirement. Allowing for negative roots could lead to more solutions, possibly with both types of twists, as recently found in \cite{Ferrero:2021etw,Couzens:2021cpk}. }
Near the endpoints of this interval, the metric becomes (we denote both roots collectively as $y_i$)
\begin{equation}
\begin{split}
	\d s²_Σ = \frac{1}{q^\prime(y_i)} \left(\frac{\d y²}{y-y_i}+\frac{\left(q^\prime(y_i)\right)²\left(y-y_i\right)}{r(y_i)}\d z²\right)
	= \frac{1}{\left|q^\prime(y_i)\right|} \left(d x² + x²\frac{\left(q^\prime(y_i)\right)²}{4r(y_i)}\d z²\right),
\end{split}
\end{equation}
where in the second step we have changed coordinates to a ``near the pole'' coordinate $x$ defined by $y-y_i=\pm x²/4$. We demand that the $z$ coordinate is periodic with period $Δz$. This requires the following conditions
\begin{equation}\label{eq:deficit0}
	\frac{q^\prime(y_i) Δz}{2\sqrt{r(y_i)}} \overset{!}{=} \pm\frac{2π}{n_\pm}\,,
\end{equation}
where the upper sign corresponds to $y=y_1$ where $q^\prime(y_1)>0$, and the lower sign to $y=y_2$ where $q^\prime(y_2)<0$. This corresponds to a metric on the spindle that is regular everywhere except at the endpoints of the interval $y \in [ y_1, y_2 ]$ where there is a conical singularity with a deficit angle $α=2π(1-n_\pm^{-1})$.
Changing from $y$ to $\tilde{y}$ defined by $y \mapsto \left(\tilde{y}³-f_3\right)/f_i$, we can solve \fref{eq:deficit0} to find an implicit equation for the roots $y_i$
\begin{equation}
	\tilde{y}_{1,2}³ = f_3 + \tilde{y}_{1,2}\left(\frac{3^{7/3}}{2^{11/3}m²} \pm \frac{3π}{16 m³ n_\pm Δz}\right)\,.
\end{equation}
We can now compute the Euler number for the metric in \fref{eq:metricspindle}. This is given by the integral of the Ricci scalar over the manifold
\begin{equation}
	χ(Σ) = \frac{1}{4π} \int_Σ  \d y\, \d z  \sqrt{g} R_Σ
	= \frac{1}{4π} \int_Σ \d y\, \d z \left(\frac{q r^\prime - q^\prime r }{r^{3/2}}\right)^\prime = \left. \frac{Δz}{4π} \frac{q r^\prime - q^\prime r }{r^{3/2}} \right\rvert_{y=y_2}^{y=y_3}
	= \left(\frac{1}{n_-}+ \frac{1}{n_+}\right)\,,
\end{equation}
which is indeed the right result for the spindle. 
Let us now compute the flux of the R-charge gauge field on the spindle
\begin{equation}\label{eq:R-flux-4d}
	\frac{g}{2π}\int_Σ F^3 = \left(\frac{1}{n_+}+\frac{1}{n_-}\right)\,.
\end{equation}
Remarkably, this is equal to the Euler number of the spindle.
The present situation resembles the ``topological twist'' that happens when compactifying on a Riemann surface of constant curvature. However, the local curvature on a spindle is not constant, therefore, the twist is like a topological twist, but only topologically. Hence this was referred to in \cite{Ferrero:2021wvk} as a topologically topological twist. In contrast, the situation that we will find in \fref{sec:solution1} is usually called an ``anti twist''.

\subsection{Free energy on $S^3$}
We have found a solution of the form $\mathrm{AdS}_4 \times Σ$ in 6d F(4) gauged supergravity.
Since this is a consistent truncation of mIIA supergravity, the solution can be uplifted to 10d. The full 10d solution corresponding to this should be thought of as an interpolating solution between $\mathrm{AdS}_6 \times S^4$ and $\mathrm{AdS}_4 \times Σ \times S^4$. The $\mathrm{AdS}_6 \times S^4$ solution is dual to a 5d $\mathcal{N}=1$ SCFT. So we expect the $\mathrm{AdS}_4 \times Σ \times S^4$ solution to be dual to the 3d SCFT obtained by compactifying the 5d SCFT on $Σ$. The free energy can be computed holographically to get
\begin{equation}
	F_{S^3} = \frac{π L²_{\mathrm{AdS}_4}}{2G^\mathrm{\s{N}}_{4d}} = \frac{2π L²_{\mathrm{AdS}_4}}{9G^\mathrm{\s{N}}_{6d}} \int \d y\,\d z\, \frac{w^2}{\sqrt{r}}
	=\frac{2π L²_{\mathrm{AdS}_4}}{9G^\mathrm{\s{N}}_{6d}}\left(\frac{Δz\left(\tilde{y}_2-\tilde{y}_1\right)}{4\cdot 6^{1/3}m²} 
	- \frac{π\left(n_+ \tilde{y}_2 + n_- \tilde{y}_1\right)}{12\cdot 6^{2/3}m³n_+ n_-} \right)\,.
\end{equation}
Since the function $q$ is a polynomial of order 8, the explicit form of the roots $\tilde{y}_{1,2}$ is difficult to obtain. 
Instead, we expand the free energy as a perturbation series in the total magnetic charge on the spindle $Q$, which is defined as
\begin{equation}\label{eq:Q-4d}
	Q = \frac{g}{2π}\int_Σ F^I = \frac{16 Δz f_i m^3}{3π}\left(\frac{1}{\tilde{y}_1}-\frac{1}{\tilde{y}_2}\right).
\end{equation}
Expanded around $Q=0$, the free energy is\footnote{This perturbative expansion around an integer $Q$ is just a trick that we have used due to the difficulty of finding the analytic form of the roots $\tilde{y}_{1,2}$, which are solutions to an order $8$ polynomial equation. If one manages to find these analytical expressions, this trick can be avoided altogether and the computation can be done exactly. We have checked numerically that the free energy matches the extremized entropy functional for arbitrary $Q$, which justifies the trick in this case.}
\begin{equation}\label{eq:FS3-pert}
\begin{split}
	F_{S^3} &= \frac{2π L²_{\mathrm{AdS}_4}}{9G^\mathrm{\s{N}}_{6d}} \cdot\\
	&\quad \left[-\frac{π\left(n_+ + n_-\right)^3}{96 m^4 n_+ n_- \left(n_+² - n_+ n_- + n_-²\right)} + \frac{π n_+n_-\left(n_++n_-\right)\left(n_+-2n_-\right)\left(2n_+-n_-\right)Q²}{192 m^4\left(n_+² - n_+ n_- + n_-²\right)²}\right]\\
	&\quad + \mathcal{O}\left(Q^4\right)\,.
\end{split}
\end{equation}
We want to compare this to the free energy of the 3d SCFT dual to this solution. This can be obtained by computing the logarithm of the inverse of the partition function of the 5d SCFT placed on $S^3 \times \Sigma$, and then taking the large $N$ limit. However, the same result can also be obtained holographically by using the technology of ``gravitational blocks'' introduced in \cite{Hosseini:2019iad}. This involves extremizing an entropy functional that is constructed by gluing gravitational blocks. The gravitational blocks are constructed from the prepotential which in this case is $\mathcal{F}\left(X_i\right) = \left(X_1 X_2\right)^{3/2}$, and is given by $\mathcal{B}\left(X_i\right) = \mathcal{F}\left(X_i\right)/ϵ$. We then define the entropy functional to be
\begin{equation}\label{eq:I-ads4}
	I = \frac{π²L_{\mathrm{AdS}_6}^4}{3G^\mathrm{\s{N}}_{6d}}\frac{8}{27}\left[ \mathcal{B}\left(X^+_i\right) - \mathcal{B}\left(X^-_i\right)+ λ\left(Δ_1 + Δ_2 - 2\right) \right]\,,
\end{equation}
where
\begin{equation}
\begin{split}
	X^\pm_1 = Δ_1 \mp \frac{ϵ}{2n_\pm} \pm \frac{s\,ϵ}{4} \,, \quad
	X^\pm_2 = Δ_2 \mp \frac{ϵ}{2n_\pm} \mp \frac{s\,ϵ}{4} \,.
\end{split}
\end{equation}
$Δ_i$ and $ϵ$ are chemical potentials conjugated to the electric charges and 
the rotational symmetry of the spindle respectively.\footnote{Our entropy functional can related to that of \cite{Faedo:2021nub} by taking their  $r_i = 1$, $n_i = (1\pm z)(n_- + n_+)/(n_- n_+)$, and redefining $z$ in terms of $s$.}
$λ$ is a Lagrange multiplier that enforces the constraint on the chemical potentials. The $\pm$ index on $X_i$ corresponds to the values at the north and the south poles of the hemispheres which are glued together, and the relative minus sign between the blocks corresponds to the ``A-gluing'' in \cite{Hosseini:2019iad}. The coefficient involving $L_{\mathrm{AdS}_6}$ and $G^\mathrm{\s{N}}_{6d}$ in \fref{eq:I-ads4} is the free energy on $S^5$, and the factor of $8/27$ comes from the coefficient in front of $\mathcal{F}$, as well as our normalization of $ϵ$. This is an off-shell expression which needs to be extremized with respect to the chemical potentials. $s$ is a continuous flavor charge and should correspond to $Q$ in \fref{eq:Q-4d}. We can perform the extremization
\begin{equation}
	∂_{Δ_i}I = ∂_{ϵ}I = 0,
\end{equation}
perturbatively around $s=0$. Upon extremization, we find that the entropy functional evaluated at the saddle point is
\begin{equation}
\begin{split}
	I_* &=\frac{8π²L_{\mathrm{AdS}_6}^4}{81G^\mathrm{\s{N}}_{6d}}\left[ -\frac{3\left(n_+ + n_-\right)^3}{8 n_+ n_- \left(n_+² - n_+ n_- + n_-²\right)} + \frac{3n_+n_-\left(n_++n_-\right)\left(n_+-2n_-\right)\left(2n_+-n_-\right)s²}{16\left(n_+² - n_+ n_- + n_-²\right)²} \right]\\
	&\quad + \mathcal{O}\left(s^4\right)\,.
\end{split}
\end{equation}
Comparing with \fref{eq:FS3-pert}, we see that they match with the identification $s=Q$.\footnote{Where we have used the identification $L²_{\mathrm{AdS}_4}=16m^4 L²_{\mathrm{AdS}_6}$.} The result can be checked to arbitrary order in the perturbation series. This shows that the free energy of the solution indeed matches the expectation from field theory and lends support to the duality that we suspected. We have performed further numerical checks for arbitrary charge $Q$ and confirmed that the entropies are indeed equal.
\section{Supersymmetric black hole solutions}\label{sec:ads2}
\subsection{Supersymmetry equations}
We will now shift attention to solutions with a different topology, namely $\mathrm{AdS}_2 \times Σ \times Σ_\mathfrak{g}$, where $Σ_\mathfrak{g}$ is a smooth Riemann surface of genus $\mathfrak{g}$.\footnote{Normalizing the metric such that $R_{mn}=κg_{mn}$, volume of the Riemann surface is $\mathrm{vol}_{Σ_\mathfrak{g}}=4π\lvert \mathfrak{g}-1 \rvert$ for $\mathfrak{g}\neq 1$, and $\mathrm{vol}_{Σ_\mathfrak{g}}=2π$ for $\mathfrak{g}=1$. For $κ=-1$, the metric is locally $H^2$, and can be quotiented to obtain a constant curvature Riemann surface with $\mathfrak{g}>1$.}
This can be thought of as the spacetime near the horizon of a black hole with the horizon geometry $Σ \times Σ_\mathfrak{g}$.
Let us consider the following metric
\begin{equation}\label{eq:ads2ansatz}
	\d s² = w(y) \left[\frac{4}{9}\d s²_{\mathrm{AdS}_2} - \frac{\d y²}{q(y)} - \frac{q(y)}{r(y)}\d z² \right] - w_1(y) \d s²_{Σ_\mathfrak{g}} \,.
\end{equation}
We assume that the scalar field $ϕ_3(y)$ depends only on the coordinate $y$. This time, we turn on a non-zero two-form, and
 pick the non-zero components of the gauge fields (where again, the index $i$ labels the gauge field from the vector multiplet) to be
\[A^3=A^3_z(y)\d z + A^3_ϕ(θ)\d ϕ\,,\quad
 A^i = A^i_z(y)\d z + A^i_ϕ(θ)\d ϕ\,,\quad 
 B=B_{tr}(y)\d t ∧ \d r \,,\]
where $θ,ϕ$ are along the Riemann surface.
It is consistent to choose a two-form with zero field strength \ie, $H_{μνρ}=0$, and we will make this simplifying assumption.
With this, the only non-trivial equations of motion for the gauge field and the two-form are given by\footnote{As before, for brevity of presentation, we will not write the $y$ dependence of the fields explicitly.}
\begin{equation}
\begin{split}
	&∂_y \left(\sqrt{-g}\,e^{-2σ} \mathcal{N}_{ΛΣ}F^{Σyz}\right) = 
	∂_θ \left(\sqrt{-g}\,e^{-2σ} \mathcal{N}_{ΛΣ}F^{Σθϕ}\right) = 0\,,\\
	&\frac{m²}{4}\sqrt{-g}\,e^{-2σ}B_{tr}g^{tt}g^{rr} + \frac{1}{8}η_{ΛΣ} F^Λ_{yz} F^Σ_{θϕ} = 0\,.
\end{split}
\end{equation}
For the U(1) gauge fields, this implies 
\begin{equation}\label{eq:F3FI}
	\begin{split}
		F^3 &= \frac{e^{2σ}}{w_1}\left(f_3\cosh2ϕ_3+f_i\sinh 2ϕ_3\right) \mathrm{vol}_Σ + \tilde{f}_3 \mathrm{vol}_{Σ_\mathfrak{g}} \,,\\
		F^I &= \frac{e^{2σ}}{w_1}\left(f_3\sinh 2ϕ_3+f_i\cosh2ϕ_3\right) \mathrm{vol}_Σ + \tilde{f}_i \mathrm{vol}_{Σ_\mathfrak{g}} \,,
	\end{split}
\end{equation}
where $f_3, f_i, \tilde{f}_3, \tilde{f}_i$ are constants.
The equation of motion for the two-form is purely algebraic and can be solved to give
\begin{equation}\label{eq:Bsol0}
	B_{tr} = \frac{2 R^4}{9m²}\,,
\end{equation}
where $R$ is a constant.
Additionally, $H_{μνρ}=0$ fixes the dilaton $σ$ in terms of the scalar field $ϕ_3$, and the warp factor on the Riemann surface $w_1$
\begin{equation}\label{eq:constraint0}
	e^{-4σ}w_1²R^4 = \left( f_3 \tilde{f}_3 - f_i \tilde{f}_i \right) \cosh 2ϕ_3 + 
	\left( f_i \tilde{f}_3 - f_3 \tilde{f}_i \right) \sinh 2ϕ_3\, \coloneqq f\!\! \cdot \!\! \tilde{f}\,.
\end{equation}
We choose the following representation for the $8\times8$ gamma matrices\footnote{As before, numerical indices are frame indices. $0,1$ lie along AdS$_2$, $2,3$ lie along the spindle $Σ$, and $4,5$ lie along the Riemann surface $Σ_\mathfrak{g}$.}
\begin{equation}
	Γ^{0,1} = γ^{0,1} \otimes σ_3 \otimes σ_3, 
    Γ^2 = \mathbbm{1} \otimes iσ_2 \otimes σ_3, 
    Γ^3 = \mathbbm{1} \otimes iσ_1 \otimes σ_3, 
	Γ^4 = \mathbbm{1} \otimes \mathbbm{1} \otimes iσ_2, 
    Γ^5 = \mathbbm{1} \otimes \mathbbm{1} \otimes iσ_1\,.
\end{equation}
In this section, we will consider compactifications on a negatively curved Riemann surface \ie, $κ=-1$.
For the 6d spinor, we choose $Γ^{45} ϵ_A = i ϵ_A$, and for the remaining 4d part of the spinor, we proceed similar to \fref{sec:ads4}. In addition, we choose to work in a gauge where the spinor is independent of the coordinate $z$ as well as the coordinates on the Riemann surface.
The supersymmetric solution can now be obtained by imposing that the fermionic variations in \fref{eq:bpseqs0} vanish. Similar to \fref{sec:ads4}, we define an approprite combination of the components of the spinor on the spindle, which we call $ξ$. In terms of this, we obtain the following set of BPS equations
\begin{equation}\label{eq:bps-ads2}
\begin{split}
	δχ &: \left(\frac{9me^{-σ}}{32w}B_{tr} - \frac{e^σa_1}{8w w_1} \right) σ_3 ξ - \left(\frac{e^{-σ}a_2}{8w_1} 
	+ \frac{a_5}{4} \right) ξ - \frac{1}{2}\sqrt{\frac{q}{w}}σ^\prime σ_2 ξ \overset{!}{=} 0\,,\\
	δλ &: \frac{e^σ a_4}{w w_1} σ_3 ξ + \left( \frac{e^{-σ} a_3}{w_1}
	+ 2 g e^σ \sinh ϕ_3\right) ξ + \sqrt{\frac{q}{w}}ϕ_3^\prime σ_2 ξ \overset{!}{=} 0\,,\\
	δψ_{Aa} &:  \left(1 - \frac{e^σ a_1}{6\sqrt{w}w_1} - \frac{9me^{-σ}}{8\sqrt{w}}B_{tr}\right) ξ - \left(\frac{e^{-σ} a_2}{6w_1} + \frac{a_6}{3} \right)\sqrt{w} σ_3 ξ - \frac{iw^\prime \sqrt{q}}{3w}σ_1 ξ \overset{!}{=} 0 \,,\\
	δψ_{A2} &: i \left(\frac{3 e^σ a_1}{8 w w_1} + \frac{9 m e^{-σ }}{32 w}B_{tr}\right) σ_1 ξ - \left(\frac{e^{-σ}a_2}{8 w_1} + \frac{a_6}{4} \right) σ_2 ξ + \sqrt{\frac{q}{w}} ξ^\prime \overset{!}{=}0\,,\\
	δψ_{A3} &: -i \left(\frac{3 e^σ a_1}{8 w w_1} + \frac{9 m e^{-σ }}{32 w}B_{tr}\right) σ_2 ξ - \left(\frac{e^{-σ}a_2}{8 w_1} + \frac{a_6}{4}\right) σ_1 ξ - \frac{ig\sqrt{r}}{2\sqrt{wq}}A^3_z ξ \\
    &\quad + \frac{i\sqrt{r}}{2w}\left(\sqrt{\frac{wq}{r}}\right)^\prime σ_3 ξ
	\overset{!}{=}0 \,,\\
	δψ_{A4} &: \left(-\frac{e^σ a_1}{8 w w_1} + \frac{9 m e^{-σ }}{32 w}B_{tr}\right) σ_3 ξ + \left(\frac{3e^{-σ}a_2}{8 w_1} - \frac{a_6}{4}\right) ξ + \frac{w_1^\prime \sqrt{q}}{4w_1 \sqrt{w}}σ_2 ξ \overset{!}{=}0 \,,
\end{split}
\end{equation}
where we have defined the following combinations
\begin{equation}
\begin{split}
    a_1 &= \left(f_3 \cosh ϕ_3 + f_i \sinh ϕ_3\right)\,,\quad
    a_2 = \left(\tilde{f}_3\cosh ϕ_3 - \tilde{f}_i \sinh ϕ_3 \right)\,,\\
    a_3 &= \left(\tilde{f}_i \cosh ϕ_3 - \tilde{f}_3 \sinh ϕ_3 \right) \,,\quad
    a_4 = \left(f_3 \sinh ϕ_3 + f_i \cosh ϕ_3\right)\,\\
    a_5 &= \left(g \cosh ϕ_3 e^σ -3m e^{-3σ}\right)\,,\quad
    a_6 = \left( g \cosh ϕ_3 e^σ+me^{-3σ} \right)\,.
\end{split}
\end{equation}
The remaining variation $δψ_{A5}\overset{!}{=}0$ gives the same condition as $δψ_{A4} \overset{!}{=}0$, along with the condition that the R-symmetry gauge field along $Σ_\mathfrak{g}$ cancels the spin connection. This is the usual topological twisting condition when compactifying on a Riemann surface\footnote{Recall that we have chosen $κ=-1$ and $g=3m$.}
\begin{eqnarray}\label{eq:twistcondition}
	\tilde{f}_3 + \frac{κ}{2g} = 0 \Rightarrow \tilde{f}_3 = \frac{1}{6m}.
\end{eqnarray}
Note that all of the BPS equations above should be supplemented with the constraint in \fref{eq:constraint0} and the value of the two-form in \fref{eq:Bsol0}. For the sake of brevity, we don't write them explicitly.

\subsection{Solution with a constant scalar}\label{sec:solution1}
We will now solve the BPS equations. To simplify the system of equations, we  further restrict to a family of solutions in which the fluxes on the Riemann surface are identified: $\tilde{f}_3 = \tilde{f}_i$. 
With this choice, \fref{eq:constraint0} becomes
\begin{equation}\label{eq:constraint1}
	e^{-4σ}w_1²R^4 = \tilde{f}_i \left(f_3-f_i\right)e^{-2ϕ_3}\,.
\end{equation}
Motivated by this, we pick a particular value of $R$ that simplifies the equations significantly:
\begin{equation}\label{eq:Rspecial}
	R^4 = 144 m^4 \tilde{f}_i \left(f_3 - f_i\right)\,.
\end{equation}
This fixes the two-form $B_{tr}$, which we take to be real and therefore restrict ourselves to the family of fluxes which have $f_3>f_i$.
As a further simplification, we assume $ϕ_3^\prime=0$. We will drop this assumption in the next subsection and construct a solution for arbitrary $ϕ_3$.
With these simplifications, we can solve the BPS equations to find a simple solution
\begin{equation}
\begin{split}
	q_{1,2} &= \frac{1}{w^\prime} \left[\pm\left(3w - 24\cdot 3^{1/8}m²f_i \sqrt{w}\right)
	+ 4\cdot 3^{3/8}m w^{3/2}\right]\,,\\
	w_1 &= \frac{1}{4\cdot 3^{3/4}m²}\,,\quad 
	e^{2ϕ_3} = \frac{1}{3}\,,\quad 
	r = r_0 \frac{w^3}{\left(w^\prime\right)²}\,,\\
	A^3_z &= \frac{2\cdot 3^{3/8}}{\sqrt{w r_0}}\left(\sqrt{w} - 16 \cdot 3^{1/8}m²f_i\right)\,, \quad
	n = n_0 \sqrt{\frac{w^\prime}{w}}\,,
\end{split}
\end{equation}
with $f_3 = 2f_i$.
$r_0$ is again an unphysical parameter that can be absorbed in a coordinate redefinition of $z$. It does not appear in any physical quantity and so we will not bother to specify it here.

\subsubsection{Regularity of the solution}\label{sec:regularity-solution-1}
We have obtained a solution of the form $\mathrm{AdS}_2\times Σ \times Σ_\mathfrak{g}$ where the scalar field $ϕ_3$ as well as the size of $Σ_\mathfrak{g}$ is constant, while the metric factors $q$ and $r$ depend on the warp factor $w$.
Let us now examine the function $q$ to ensure that the metric on $Σ$ is smooth everywhere except at the poles. 
We choose a parametrization of the function $w$ as $w(y)=y²$. With this explicit choice, the metric coefficient $r$ is positive, and $q$ is a reduced quartic polynomial
\begin{equation}
	q = q_1 q_2 = 4\cdot 3^{3/4}m²y^4 - \frac{9y²}{4} + 36\cdot 3^{1/8}m²f_i \,y -144\cdot 3^{1/4}m^4 f_i²\,.
\end{equation}
For $f_i>0$, it has four real roots\footnote{We have restricted ourselves to positive roots as discussed in \fref{fn:positiveroots}.} given by
\begin{equation}
\begin{split}
	y_{1,2} = \frac{3^{1/8}}{8m} \left( -√3 \mp \sqrt{3 + 128√3m³f_i} \right),\quad
	y_{3,4} = \frac{3^{1/8}}{8m} \left( √3 \mp \sqrt{3 - 128√3m³f_i} \right)\,,
\end{split}
\end{equation}
where the order of the indices correspond to the upper and lower signs respectively.
The function $q$ is positive when $y$ lies in the closed interval $y_2 \leq y \leq y_3$, and both roots are positive when $0 < 128 m³ f_i < \sqrt{3}$. In fact, $y_2$ comes from $q_1=0$, while $y_3$ comes from $q_2=0$.
\Fref{eq:deficit0} can now be solved to find the period of $z$ and to determine the flux
\begin{equation}\label{eq:415}
	f_i = \frac{√3}{128m³}\left(\frac{n_-² - n_+²}{n_-² + n_+²}\right)\,, \quad
	Δz = \frac{πr_0}{3\sqrt{2}3^{3/8}m}\sqrt{\frac{1}{n_-²}+ \frac{1}{n_+²}}\,.
\end{equation}
As a consistency check, we can again compute the Euler number for the metric in \fref{eq:metricspindle}. This is given by the integral of the Ricci scalar
\begin{equation}
	χ(Σ) = \frac{1}{4π} \int_Σ  \d y\, \d z  \sqrt{g} R
	= \left. \frac{Δz}{4π} \frac{q r^\prime - q^\prime r }{r^{3/2}} \right\rvert_{y=y_2}^{y=y_3}
	= \left(\frac{1}{n_-}+ \frac{1}{n_+}\right)\,,
\end{equation}
which is indeed the right result for the spindle.
Let us now evaluate the total R-charge on the spindle. This integral receives contributions only from the endpoints of the interval to give
\begin{equation}\label{eq:R-flux_solution-1}
	\frac{g}{2π}\int_Σ F^3 = \frac{g}{2π}Δz \left[A^3_z(y_3)-A^3_z(y_2)\right] 
	= \left(\frac{1}{n_+}-\frac{1}{n_-}\right)\,.
\end{equation}
As alluded to in the previous section, this is not equal to the Euler number, but rather corresponds to an ``anti twist''
Finally, we can compute the Bekenstein-Hawking entropy of this black hole 
\begin{equation}\label{eq:area-solution-1}
	S_\mathrm{\s{BH}} 
	= \frac{\mathrm{Area}_{Σ \times Σ_\mathfrak{g}}}{4G^\mathrm{\s{N}}_{6d}} 
	= \frac{\mathrm{vol}_{Σ_\mathfrak{g}}}{4G^\mathrm{\s{N}}_{6d}}\int \d y \d z \frac{w w_1}{\sqrt{r}}
	= -\frac{π\left(n_- + n_+ - √2\sqrt{n_-² + n_+²}\right)}{48 √3 m^4 n_- n_+}\frac{1}{4G^\mathrm{\s{N}}_{4d}}\,,
\end{equation}
where in the last step we have used $G^\mathrm{\s{N}}_{4d}=G^\mathrm{\s{N}}_{6d}/\mathrm{vol}_{Σ_\mathfrak{g}}$.
Note that the expression within the parenthesis is always negative, and so the area has the correct sign.


\subsection{Solution with non-constant scalar}\label{sec:solution2}
Let us now drop the assumption that $ϕ_3^\prime=0$ and look for a solution with a non-constant scalar. We will still restrict ourselves to the family of fluxes where
\begin{equation}
	\tilde{f}_3 = \tilde{f}_i, \quad
	R^4 = 144 m^4 \tilde{f}_i \left(f_3 - f_i\right)\,.
\end{equation}
The BPS equations \eqref{eq:bps-ads2} now have the following solution
\begin{equation}\label{eq:solution-non-constant-scalar}
\begin{split}
	q_{1,2} &= \frac{e^{-2ϕ_3}}{8\left(f_3 - 2f_i\right)ϕ_3^\prime}
	\left[\pm\left( 3\left(f_3-f_i\right) - 6 e^{2ϕ_3}\left(2f_3-f_i\right) + 9 e^{4ϕ_3} \left(f_3+f_i\right) \right) \right. \\
	& \hskip 90pt \left. - 128 e^{3ϕ_3}m³\left(f_3-2f_i\right)² \right]\,,\\
	w_1 &= \frac{e^{-ϕ_3/2}}{12m²}\,,\quad
	w = 256 e^{7ϕ_3/2}m^4 \left(\frac{f_3-2f_i}{3e^{2ϕ_3}-1}\right)²\,,\quad
	r = r_0² \frac{e^{2ϕ_3}}{\left(ϕ_3^\prime\right)²}\,,\\
	A^3_z &= \frac{2m²e^{-2ϕ_3}}{r_0} \left[ \left(3e^{4ϕ_3}+4e^{2ϕ_3}-3\right)f_3 + 
	\left(3e^{4ϕ_3}-2e^{2ϕ_3}+3\right)f_i \right]\,,\\
	n &= n_0 e^{3ϕ_3/8}  \sqrt{\frac{ϕ_3^\prime}{3e^{2ϕ_3}-1}}\,.
\end{split}
\end{equation}
We have now obtained a solution in which the scalar field is not constant. The dilaton as well as the metric factors are determined in terms of the arbitrary scalar $ϕ_3$. The solution has two free parameters $f_3$ and $f_i$ corresponding to the fluxes, in contrast to the solution with constant scalars which had only one free parameter $f_i$. So we expect this solution to reduce to the one in the previous subsection under a specific choice of fluxes. To see this, let us trade $ϕ_3$ for a new function $A(y)$ which we define by the following relation
\begin{equation}
	e^{2ϕ_3} = \frac{f_3-2f_i + A}{3A}\,,
\end{equation}
and rescale the arbitrary constant $r_0$ to $r_0 = \tilde{r}_0 \left(f_3 - 2f_i\right)$.
The solution in \fref{eq:solution-non-constant-scalar} can be rewritten in terms of $A$ as follows
\begin{equation}\label{eq:solution-non-constant-scalar-A}
\begin{split}
	w_1 &= \frac{1}{4\cdot 3^{3/4}m^2}\left(\frac{A}{f_3-2f_i+A}\right)^{1/4}\,,\quad
	w = \frac{256 m^4 A^{1/4}}{3\cdot 3^{3/4}}\left(f_3-2f_i+A\right)^{7/4}\,,\\
	r &= \frac{4A\tilde{r}_0²}{3\left(A^\prime\right)^2}\left(f_3-2f_i+A\right)^3\,,\quad
	n = \frac{n_0 \sqrt{A^\prime}}{√2 \left[27A^3\left(f_3-2f_i+A\right)^5\right]^{1/16}}\,,\\
	q_{1,2} &= \frac{\mp 9 \left(f_3+f_i-2A\right) + 128 √3m³\sqrt{A}\left(f_3-2f_i+A\right)^{3/2}}{12A^\prime}\,.
\end{split}
\end{equation}
As before, the constraint in \fref{eq:constraint1} determines $σ$. We have checked explicitly that this solution in terms of $A$ is a solution to the BPS equations. It is now easy to see that imposing $f_3 = 2f_i$ takes us back to the solution with a constant scalar in \fref{sec:solution1}.

\subsubsection{Regularity of the solution}
We have obtained a solution with the scalars and the metric factors depending on a single arbitrary function $A$. To analyse the structure of the metric, we will pick this function to be $A(y)=y$. We will further define fluxes $g_1,g_2$ as the following linear combinations of $f_3,f_i$
\begin{equation}
	g_1 = f_3 + f_i\,,\quad
	g_2 = f_3 - 2f_i\,.
\end{equation}
In terms of these fluxes, setting $g_2=0$ takes us back to the solution in \fref{sec:solution1}. With this choice,
\begin{equation}
	q = \frac{1024}{3}m^6 y\left(y+g_2\right)³ - \frac{9y²}{4} + \frac{9 g_1y}{4} - \frac{9g_1²}{16}\,,
\end{equation}
which is again a quartic equation (but now including a $y³$ term as well) with a positive coefficient for the leading term. This has four roots, with constraints on $g_1, g_2$ for all the roots to be real. In particular, we are interested in the cases where the middle two roots (we will call them $y_2,y_3$ like before) are positive. The interval $0 < y_2 \leq y \leq y_3$ then corresponds to a positive $q$. The metric coefficient $r$ on the other hand is positive for $f_3 \geq 2f_i$. It is much more difficult to find the roots analytically in our present solution.
Therefore, we have repeated the regularity analysis of \fref{sec:regularity-solution-1} by picking numerical values of the fluxes, and checked that it correctly reproduces the Euler character of the spindle, as well the total flux in \fref{eq:R-flux_solution-1}.

Furthermore, to get an analytic handle on the entropy, we have performed a perturbative expansion as a series in $g_2$ around $g_2=0$ and checked against numerical results. We will briefly present this here. The strategy is the following: the two equations \eqref{eq:deficit0} describing the deficit angles determine the flux $g_1$ as well as the periodicity $Δz$ in terms of $g_2,y_2,y_3$, and positive coprime integers $n_\pm$. This leaves $g_2$ undetermined. Further, $q\left(y_2\right)=q\left(y_3\right)=0$ determines $y_{2,3}$ in terms of $g_2, n_\pm$. Since $g_2=0$ corresponds to the constant scalar solution of \fref{sec:solution1}, it is natural to expand all quantities as a perturbation series in $g_2$. The roots $y_{2,3}$ expanded in $g_2$ read
\begin{equation}
\begin{split}
	y_2 &= -\frac{3√3}{128m³}\left(1-\frac{√2n_-}{\sqrt{n_+²+n_-²}}\right) -\frac{3g_2}{4} 
	+ \frac{4√2m³n_- g_2²}{\sqrt{n_+²+n_-²}} 
	+ \mathcal{O}\left(g_2^3\right)\,,\\
	y_3 &= \frac{3√3}{128m³}\left(1-\frac{√2n_+}{\sqrt{n_+²+n_-²}}\right) -\frac{3g_2}{4} 
	- \frac{4√2m³n_+g_2²}{\sqrt{n_+²+n_-²}}  + \mathcal{O}\left(g_2^3\right)\,.
\end{split}
\end{equation}
We can now also compute the area as an expansion in $g_2$. The relevant physical quantity corresponding to $g_2$ is the magnetic charge on the spindle
\begin{equation}
	Q \coloneqq \frac{g}{2π}\int_Σ F^I\,.
\end{equation}
In the absence of $g_2$, the flux on the spindle is $Q_0 = \left(n_+-n_-\right)/\left(4n_- n_+\right)$. Subtracting this constant flux from $Q$, we define $\tilde{Q} \coloneqq Q - Q_0$, and rewrite $g_2$ as an expansion in this parameter to get
\begin{equation}
	g_2 = \frac{√6 n_+ n_- \tilde{Q}}{16m³\sqrt{n_+²+n_-²}} 
	+ \frac{n_+² n_-² \left(n_+ + n_- +√2\sqrt{n_+²+n_-²} \right)\tilde{Q}²}{2√6 m³ \left(n_+²-n_-²\right)\sqrt{n_+² + n_-²}} 
	+ \mathcal{O}\left(\tilde{Q}^3\right) \,.
\end{equation}
We can now compute the area as an expansion in $\tilde{Q}$,
\begin{equation}\label{eq:area-ads2-2}
\begin{split}
	S_\mathrm{\s{BH}} 
	&= \frac{\mathrm{Area}_{Σ \times Σ_\mathfrak{g}}}{4G^\mathrm{\s{N}}_{6d}}
	= \frac{\mathrm{vol}_{Σ_\mathfrak{g}}}{4G^\mathrm{\s{N}}_{6d}}\int \d y\, \d z \frac{w w_1}{\sqrt{r}}\\
	& = \frac{1}{4G^\mathrm{\s{N}}_{4d}} 
	\left[ - \frac{π\left(n_- + n_+ - √2\sqrt{n_-² + n_+²}\right)}{48 √3 m^4 n_- n_+}
	- \frac{π n_+ n_- \tilde{Q}²}{72√6m^4\sqrt{n_-² + n_+²}} \right]
	+ \mathcal{O}\left(\tilde{Q}^3\right) \,,
\end{split}
\end{equation}
where in the last line we have used $G^\mathrm{\s{N}}_{4d}=G^\mathrm{\s{N}}_{6d}/\mathrm{vol}_{Σ_\mathfrak{g}}$.
As expected, this indeed reduces to \fref{eq:area-solution-1} for $\tilde{Q}=0$.

\subsection{Subtruncation to 4d gauged $T^3$ supergravity}
Let us now interpret the $\mathrm{AdS}_2 \times Σ \times Σ_\mathfrak{g}$ solutions 
that we have found in sections \ref{sec:solution1} and \ref{sec:solution2}, in terms of solutions to the 6d theory compactified on the Riemann surface $Σ_\mathfrak{g}$. A general compactification of this form gives a four dimensional $\mathcal{N}=2$ gauged supergravity \cite{Hosseini:2020wag}, a particular subtruncation of which is the ``gauged $T^3$ model''. This theory consists of a single vector multiplet whose scalars parametrize the coset manifold SU(1,1)/U(1). The same 4d theory can also be obtained as a consistent truncation of the ``gauged STU model'', which is the maximal four dimensional $\mathcal{N}=8$ supergravity obtained from a reduction of eleven dimensional supergravity on $S^7$ \cite{deWit:1986oxb,Cvetic:1999xp}.

The truncation from 6d F(4) gauged supergravity to the 4d ``gauged $T^3$ model'' was performed in \cite{Hosseini:2020wag}, and the 6d solutions that we have found turn out to correspond to this subtruncation. To see this, we can compare properties of our solution to those presented in \cite{Hosseini:2020wag}.
The 4d fields $\left(χ_1, χ_2, σ\right)$ can be identified with combinations of the 6d fields as follows
\begin{equation}\label{eq:4dscalars}
	e^{4χ_1} = \frac{e^{4σ}}{w_1²}\,,\quad
	χ_2 = ϕ_3\,,\quad
	e^{2ϕ} = \frac{e^{-2σ}}{w_1}\,.
\end{equation}
In the $T^3$ model, these scalars should have $e^{2ϕ} = e^{2χ_1-χ_2} = 12m²$. Using \fref{eq:solution-non-constant-scalar} or \eqref{eq:solution-non-constant-scalar-A}, we see that indeed our solution reproduces this. Additionally, if we rewrite the fluxes through $Σ_\mathfrak{g}$ in terms of $s_1$ and $s_2$
\begin{equation}\label{eq:s1s2}
	2\int_{Σ_\mathfrak{g}} F^3 = s_1 + s_2\,,\quad
	2\int_{Σ_\mathfrak{g}} F^I = s_1 - s_2\,,
\end{equation}
then the $T^3$ model has $s_1 = 1/\left(3m\right), s_2 = 0$. Recalling from \fref{eq:F3FI} that $\int_{Σ_\mathfrak{g}} F^3 = \int_{Σ_\mathfrak{g}} F^I = 1/(6m)$, we see that indeed the solutions that we have found correspond to the 4d gauged $T^3$ subtruncation of the 6d theory.

Analogous to the discussion in \fref{sec:ads4}, we expect the full solution to interpolate between $\mathrm{AdS}_2 \times Σ \times Σ_\mathfrak{g} \times S^4$ and $\mathrm{AdS}_4 \times Σ_\mathfrak{g} \times S^4$. The $\mathrm{AdS}_4 \times Σ_\mathfrak{g} \times S^4$ solution is dual to a 3d SCFT \cite{Bah:2012dg}. 
So it is natural to expect that our solution is dual to this 3d theory on a spindle. To find support for this duality, we compute the Bekenstein-Hawking entropy and compare it with a holographic computation where we minimize an entropy functional obtained by gluing gravitational blocks.
To construct this off-shell entropy functional, we start from the on-shell free energy of the $\mathrm{AdS}_4 \times Σ_\mathfrak{g}$ solution, which is the free energy on $S^3$ and in the large N limit is given by $F_{S^3} = π L_{\mathrm{AdS}_4}²/G^\s{N}_4$. We then promote this to an off-shell quantity by using the prepotential of the four dimensional magnetic STU model (and identifying three of the four fields, corresponding to the $T^3$ model) $\mathcal{F} = \sqrt{X_1 X_2^3}$. Using gravitational blocks defined by $\mathcal{B}\left(X_i\right) = \mathcal{F}\left(X_i\right)/ϵ$, we construct the following entropy functional \cite{Hosseini:2019iad}
\begin{equation}\label{eq:I-T3}
	I = \frac{π L_{\mathrm{AdS}_4}²}{2G^\mathrm{\s{N}}_{4d}}\left[ \mathcal{B}\left(X^+_i\right) + \mathcal{B}\left(X^-_i\right) + λ\left(Δ_1 + 3Δ_2 - 2\right) \right],
\end{equation}
where 
\begin{equation}
\begin{split}
	X^\pm_1 = \left(Δ_1 - \frac{ϵ}{2n_\pm} \mp \frac{s\,ϵ}{2}\,\right) \,, \quad
	X^\pm_2 = \left(Δ_2 - \frac{ϵ}{2n_\pm} \pm \frac{s\,ϵ}{6}\right) \,.
\end{split}
\end{equation}
As before $Δ_i$ and $ϵ$ are chemical potentials conjugated to the electric charges and 
the rotational symmetry of the spindle
respectively, and $λ$ is a Lagrange multiplier that enforces the constraint $Δ_1 + 3 Δ_2 = 2$.\footnote{Our entropy functional can be related to \cite{Faedo:2021nub} by taking their $r_i=1/2$, $n_1 = (1+z)(n_-- n_+)/(4n_- n_+)$, $n_2 = (1-z/3)(n_--n_+)/(4n_- n_+)$ so that $n_1 + 3n_2 = (n_--n_+)/(n_- n_+)$, and redefining $z$ in terms of $s$.}
The $\pm$ index on $X_i$ refers to the north and the south pole across which the gravitational blocks are glued.
The plus sign between the blocks in \fref{eq:I-T3} corresponds to the identity gluing in \cite{Hosseini:2019iad}. To find the entropy, we now need to extremize this functional with respect to $Δ_i$ and $ϵ$. We do this perturbatively in the parameter $s$ that corresponds to the flavor charge, to find the following extremized value
\begin{equation}
\begin{split}
	I_* &= \frac{π L_{\mathrm{AdS}_4}²}{2G^\mathrm{\s{N}}_{4d}} \left[ -\frac{\left(n_- + n_+ - √2\sqrt{n_-² + n_+²}\right)}{2 n_- n_+}
	-\frac{n_+ n_- s²}{3 √2 \sqrt{n_-² + n_+²}} \right]
	+ \mathcal{O}\left(s^3\right).
\end{split}
\end{equation}
The four dimensional AdS length is determined from the scalar potential of the four dimensional theory to give $1/L_{\mathrm{AdS}_4}² = 48√3m^4$ \cite{Hosseini:2020wag}.
Remarkably this reproduces the Bekenstein-Hawking entropy in \fref{eq:area-ads2-2} with the identification $s=\tilde{Q}$. The agreement can be checked to arbitrary orders in the perturbation series. We have also repeated the exact computation numerically and we see that the entropies indeed match exactly. This lends support to our expectation about the holographic duality.

\subsection{Solution without equal fluxes}
So far we have solved the BPS equations on the locus given by $\tilde{f}_3 = \tilde{f}_i$. We will now drop this assumption, as well as keep $κ$ arbitrary, and look for new solutions. However, we restrict ourselves only to solutions with a constant scalar $ϕ_3^\prime =0$. 
Solutions to the BPS equations are obtained in a way analogous to that outlined in the previous sections. The functions appearing in the metric are
\begin{equation}
\begin{split}
	w_1 &= -\frac{κ\sqrt{f_3}\left(f_3²-f_i²\right)^{3/4}}{6√6 m² \left(f_3²-2f_i²\right)}\,,\quad
	r = \frac{r_0 w^3}{\left(w^\prime\right)²}\,,\\
	q_{1,2} &= \pm \left(κ\frac{12\cdot 2^{3/4}3^{1/4}m²\sqrt{w}\left(f_3²-2f_i²\right)}{f_3^{3/4}\left(f_3²-f_i²\right)^{1/8}w^\prime}+\frac{3w}{w^\prime} \right) + \frac{2^{5/4}3^{3/4}m f_3^{3/4} w^{3/2}}{\left(f_3²-f_i²\right)^{3/8}w^\prime} \,.
\end{split}
\end{equation}
The scalar $ϕ_3$ and the normalization of the spinor $n$ are
\begin{equation}
	e^{2ϕ_3} = \frac{f_3-f_i}{f_3+f_i}\,,\quad
	n = n_0 \sqrt{\frac{w^\prime}{w}}\,.
\end{equation}
The R-charge gauge field along the spindle is
\begin{equation}
	A^3_z = \frac{48m²κ\left(f_3²-2f_i²\right)+2^{1/4}3^{3/4}f_3^{3/4}\left(f_3²-f_i²\right)^{1/8}\sqrt{w}}{\sqrt{\left(f_3²-f_i²\right)r_0 w}}\,,
\end{equation}
and the fluxes $\tilde{f}_{i}$ are related to $f_{3,i}$ by
\begin{equation}
	\tilde{f}_i = -\frac{κf_3 f_i}{2g\left(f_3²-2f_i²\right)}\,.
\end{equation}
This solution also turns out to have an interpretation in terms of the 6d F(4) gauged supergravity compactified on a Riemann surface. The particular subtruncation this corresponds to, is minimal supergravity in 4d. Its properties were discussed in \cite{Hosseini:2020wag}, and we can check that the explicit solution obtained here has the correct properties.
Parametrizing the fluxes in terms of $s_1, s_2$ using \fref{eq:s1s2} as before, we find
\begin{equation}\label{eq:s1s2-universal}
	s_1 + s_2 = -\frac{κ}{g}\,, \quad
	s_1 - s_2 = -\frac{κ f_3 f_i}{g\left(f_3² - 2f_i²\right)}\,.
\end{equation}
We can then compute the 4d scalars using the identification in \fref{eq:4dscalars} to find
\begin{equation}
\begin{split}
	\frac{e^{-2σ}}{w_1} &= \frac{24m²}{-κ + m \sqrt{9\left(s_1-s_2\right)²+4s_1s_2}}\,,\\
	\frac{e^{4σ}}{w_1²} &= \frac{96m³}{m\left(9\left(s_1-s_2\right)²+12s_1s_2\right) - κ\sqrt{9\left(s_1-s_2\right)²+4s_1s_2} }\,,\\
	e^{2ϕ_3} &= \frac{2s_1}{\sqrt{9\left(s_1-s_2\right)²+4s_1s_2} + 3\left(s_1-s_2\right)}\,,
\end{split} 
\end{equation}
which exactly matches the result for the 4d truncation in \cite{Hosseini:2020wag}.

\subsubsection{Regularity of the metric}
This solution is valid for a Riemann surface with arbitrary $κ$. However, to study the structure of the metric near the poles, we choose $κ=-1$, and pick the arbitrary function $w$ to be $w=y²$. The function $q$ is a reduced quartic polynomial with a positive leading coefficient
\begin{equation}
	q = \frac{3√6 f_3^{3/2} m² y^4}{\left(f_3²-f_i²\right)^{3/4}} - \frac{9y²}{4} 
	+ \frac{18\cdot 2^{3/4}\cdot 3^{1/4}\left(f_3²-2f_i²\right)m²y}{f_3^{3/4}\left(f_3²-f_i²\right)^{1/8}}
	- \frac{72 √6 \left(f_3²-2f_i²\right)²m^4}{f_3^{3/2}\left(f_3²-f_i²\right)^{1/4}}\,.
\end{equation}
This has four roots
\begin{equation}
\begin{split}
	y_{1,2} &= \frac{3^{1/4}f_1^{1/8}\left( -f_1^{1/4} \pm \sqrt{f_1^{1/2} + 64 f_2 m³} \right)}{4\cdot 2^{1/4} m \left(2f_1-f_2\right)^{3/8}}  \,,\\
	y_{3,4} &= \frac{3^{1/4}f_1^{1/8}\left( f_1^{1/4} \pm \sqrt{f_1^{1/2} - 64 f_2 m³} \right)}{4\cdot 2^{1/4} m \left(2f_1-f_2\right)^{3/8}}\,,
\end{split}
\end{equation}
where we have defined the combination of fluxes $f_1 = f_3² - f_i²$, and $f_2 = f_3² - 2 f_i²$.
For the metric to be a smooth metric on the spindle, the middle two roots must be positive, when then specify the interval for $y$ \ie, $y \in [y_2,y_3]$. The function $r$ in the metric is always positive for $r_0>0$. There are conical singularities at the ends of this interval. Parametrizing the deficit angles by coprime integers $n_\pm$, we have
 \begin{equation}
	 \frac{3^{7/4}2^{1/4}\left(2f_1-f_2\right)^{3/8}m\sqrt{f_1^{1/2} \pm 64 f_2 m³}}{\sqrt{r_0} f_1^{5/8}} = \pm \frac{2π}{n_\pm Δz}\,.
 \end{equation}
This determines one of the fluxes and the periodicity in $z$ in terms of the other flux and integers $n_\pm$. We can then compute the R-symmetry flux through the spindle, which turns out to be
\begin{equation}
	\frac{g}{2π}\int_Σ F^3 = \left(\frac{1}{n_+} - \frac{1}{n_-}\right)\,.
\end{equation}
Similar to the solutions in sections \ref{sec:solution1} and \ref{sec:solution2}, this is of the ``anti twist'' type. The Euler character of the spindle can be computed similarly and indeed gives
the correct result
\begin{equation}
	χ(Σ) = \left(\frac{1}{n_+} + \frac{1}{n_-}\right)\,.
\end{equation}
Finally, let us compute the area of the horizon of this black hole, which is given straightforwardly in terms of the single free flux parameter. It is, however, more useful to use a different parametrization. Let us rewrite the R-symmetry flux and the magnetic flux through the Riemann surface in terms of a parameter $ζ$ corresponding to the topological twist on $Σ_\mathfrak{g}$ which parametrizes the difference between the fluxes as follows
\begin{equation}
	s_1 = -\frac{κ}{6m} \left(1 + \frac{ζ}{κ}\right)\,,\quad
	s_2 = -\frac{κ}{6m} \left(1 - \frac{ζ}{κ}\right)\,.
\end{equation}
Restoring an arbitrary $κ$, the Bekenstein-Hawking entropy is
\begin{equation}\label{eq:S-ads2-universal}
\begin{split}
	S_\mathrm{\s{BH}} 
	&= \frac{\mathrm{Area}_{Σ \times Σ_\mathfrak{g}}}{4G^\mathrm{\s{N}}_{6d}}\\
	& = \frac{1}{4G^\mathrm{\s{N}}_{4d}} 
	\left[ -\frac{\left(n_+ + n_- - √2\sqrt{n_+² + n_-²}\right)π}{n_+ n_-} \cdot
	\frac{\left(\sqrt{κ²+8ζ²}-3κ\right)²}{864 m^4\sqrt{2κ\left(κ-\sqrt{κ²+8ζ²}\right)+4ζ²}} \right]\,.
\end{split}
\end{equation}
where in the last line, we have used $G^\mathrm{\s{N}}_{4d} = G^\mathrm{\s{N}}_{6d}/\mathrm{vol}_{Σ_\mathfrak{g}}$.
The second factor in the entropy is precisely $L²_{\mathrm{AdS}_4}$ for the four dimensional minimal supergravity obtained as a subtruncation of the six dimensional theory as obtained in \cite{Bah:2018lyv,Hosseini:2020wag}.
The first factor matches the entropy for $\mathrm{AdS}_2 \times Σ$ solution in 4d $\mathcal{N}=4$ gauged supergravity found in \cite{Ferrero:2020twa}, in the absence of rotation.
As a check, taking $κ=-1$ and $s_2=0$ (which corresponds to $ζ=-1$) in the above, gives $1/L²_{\mathrm{AdS}_4} = 48 √3 m^4$, and
correctly reproduces the entropy in \fref{eq:area-solution-1}. 

We can again compute the entropy holographically using gravitational blocks. The prepotential is simply given by $\mathcal{F}\left(X\right)=X²$, which defines the gravitational block $\mathcal{B}\left(X\right) = \mathcal{F}\left(X\right)/ϵ$. The constraint now fixes $Δ=1/2$ to give
\begin{equation}\label{eq:I-eq}
	I = \frac{π L_{\mathrm{AdS}_4}²}{2G^\mathrm{\s{N}}_{4d}}\left[ \mathcal{B}\left(X^+\right) + \mathcal{B}\left(X^-\right) + λ\left(4Δ - 2\right) \right],
\end{equation}
where 
\begin{equation}
	\begin{split}
		X^\pm = \left(Δ - \frac{ϵ}{2n_\pm}\right) \,.
	\end{split}
\end{equation}
Extremizing this with respect to $ϵ$ gives
\begin{equation}
	I_* = \frac{π L_{\mathrm{AdS}_4}²}{2G^\mathrm{\s{N}}_{4d}} \left[ -\frac{\left(n_+ + n_- - √2\sqrt{n_+² + n_-²}\right)}{2 n_+ n_-} \right]\,,
\end{equation}
which exactly matches the entropy in \fref{eq:S-ads2-universal} with $L²_{\mathrm{AdS}_4}$ identified as above.

\section{Discussion}\label{sec:discussion}
By solving the BPS equations in six dimensional F(4) gauged supergravity, we have found two classes of solutions: $\mathrm{AdS}_4 \times Σ$ and $\mathrm{AdS}_2 \times Σ \times Σ_\mathfrak{g}$. We conjectured that the $\mathrm{AdS}_4 \times Σ$ solution is dual to a five dimensional $\mathcal{N}=1$ SCFT on a spindle $Σ$, while the $\mathrm{AdS}_2 \times Σ \times Σ_\mathfrak{g}$ is dual to a three dimensional SCFT on $Σ$. We computed the entropy holographically by extremizing the entropy functional constructed from gravitational blocks and found that it agrees with the entropy computed from gravity. One class of our $\mathrm{AdS}_2 \times Σ \times Σ_\mathfrak{g}$ solutions corresponds to the gauged $T^3$ supergravity, while the other corresponds to minimal supergravity theory in four dimensions.

Our solutions are obtained in a six dimensional truncation of mIIA supergravity, and an uplifted solution in ten dimensions can be constructed. The four dimensional $T^3$ subtruncation of the six dimensional theory further admits an uplift in eleven dimensional supergravity. The solutions that we have found in this paper should then be expected to represent near horizon geometries of wrapped branes in ten or eleven dimensions. It would be very interesting to construct these uplifted solutions and understand the objects that they correspond to.

The solutions presented in this paper should be seen as fixed points of a flow from the supersymmetric $\mathrm{AdS}_6$ solution. Constructing the full flow is often a challenging task. While there are a few examples of full analytic flows \eg, a rotating black hole in $\mathrm{AdS}_4$ in \cite{Ferrero:2020twa,Ferrero:2021ovq,Hristov:2018spe}, it can often only be done numerically. Nevertheless, it would be interesting to construct the full flow for the present solutions to better understand the objects they describe.

Lastly, we have constructed entropy functionals by appropriately gluing gravitational blocks and we have seen that they reproduce the entropy of the gravitational solutions. Finding an explanation of these entropy functionals from field theory would be very useful.

\section*{Acknowledgement}
I am extremely grateful to Alberto Zaffaroni for collaboration in the early stages of this work, for patiently explaining to me various aspects of the problem, and for very helpful discussions. I would like to thank Eoin Ó Colgáin, Christopher Couzens, and Minwoo Suh for correspondence and useful discussions. I would also like to thank the anonymous referee for their valuable comments, which have been of immense help in improving the article. I am partially supported by the INFN and the MIUR-PRIN contract 2017CC72MK003.

\bibliographystyle{utphysmodb}
\bibliography{draft_v2.bib}
\end{document}